# Selective synthesis and crystal chemistry of candidate rare earth Kitaev materials: honeycomb and hyperhoneycomb Na$_2$PrO$_3$

Ryutaro Okuma[1,2], Kylie MacFarquharson[1], and Radu Coldea[1]

[1]Clarendon Laboratory, University of Oxford Physics Department, Parks Road, Oxford OX1 3PU, United Kingdom

[2]Institute for Solid State Physics, University of Tokyo, Kashiwa, Chiba 277-8581, Japan

**Abstract**: Rare earth oxides have attracted interest as a platform for studying frustrated magnetism arising from bond dependent anisotropic interactions. Ordered rock salt compounds Na$_2$PrO$_3$ crystallize in two polymorphs ($\alpha$- and $\beta$) comprising honeycomb and hyperhoneycomb lattices of octahedrally coordinated Pr$^{4+}$ ($4f^1$). Although possible realization of antiferromagnetic Kitaev interactions is anticipated for these phases on the basis of *ab initio* models, the air-sensitivity of the two polymorphs has hampered reliable crystal growth and physical property measurements. Here, we have succeeded in preparing powder and single crystals of both $\alpha$- and $\beta$-Na$_2$PrO$_3$ using modified synthetic procedures. Revised crystal structures for both polymorphs are obtained from refinement of untwinned single-crystal X-ray diffraction data.

## INTRODUCTION

Frustrated spin interactions between quantum spins are expected to stabilize under certain circumstances a macroscopically entangled ground state with fractional spin excitations, called a quantum spin liquid[1-3]. The Kitaev model on the honeycomb lattice is one of few exactly solvable models of a two-dimensional quantum spin liquid, in which each spin feels three types of Ising interactions depending on the bond direction[4]. After the proposal that the Kitaev model could be realized in transition metal oxides with large spin orbit coupling[5-7] numerous honeycomb magnets based on Ir$^{4+}$ ($5d^5$) and Ru$^{3+}$ ($4d^5$) have been explored[5-7] and evidence for substantial bond dependent interactions[8] and unconventional thermal transport[9] and continuum of magnetic excitations[10] have been observed. Key ingredients for the exotic interactions are spin-orbit entangled $j_{\text{eff}}=1/2$ degrees of freedom and edge-sharing octahedra of the magnetic ions. Extension of the Kitaev model to three-dimensional lattices that maintain the local-three-fold coordination of each site keeps the model exactly solvable[11] and introduces new physics[12], although realization of such lattices in real materials is limited to $\beta$-[13, 14] and $\gamma$-Li$_2$IrO$_3$[15, 16], and site-disordered $\beta$-ZnIrO$_3$[17]

**Lanthanide Kitaev Magnets** $4d$ and $5d$ transition metal oxides have been widely established as a platform for exhibiting anisotropic magnetism due to an inherent strong spin-orbit coupling. Similar bond-dependent interactions can be in principle also be realized in rare-earth $4f$ ions. This is because the ground state wave functions for $4f^1$ ions can also realize a $j_{\text{eff}} = 1/2$ Kramers ground state doublet in an octahedral cubic crystal field environment[18]. However, the large ionic radius makes it uncommon to have octahedral coordination by oxygen atoms as exemplified by 8-fold coordination in pyrochlore oxides[19]. Octahedral coordination of lanthanides is typically found for the smallest magnetic lanthanide Yb$^{3+}$ or for larger anions such as chalcogenides (S$^{2-}$, Se$^{2-}$, and Te$^{2-}$) and halides (Cl$^-$, Br$^-$, and I$^-$). Two-dimensional triangular lattice compounds with delafossite structures[20-23] and honeycomb lattice halides[24-26] of lanthanides have recently attracted attention in the context of frustrated quantum magnetism. Recent *ab-initio* calculations proposed that rare-earth ions connected by edge-sharing octahedra[27-30] could potentially realize the quite elusive antiferromagnetic Kitaev interactions, predicted theoretically to stabilize much richer phase diagrams in applied field than systems with ferromagnetic Kitaev interactions.

**Tetravalent lanthanides** Another way to make lanthanides with octahedral coordination is to use a tetravalent state, such as found in the case of binary oxides $Ln$O$_2$ with $Ln$ = Ce, Pr and Tb. Pr$^{4+}$ ($4f^1$) is of particular interest because it has a Kramers ground state doublet with an effective $j_{\text{eff}}$ = 1/2 and ionic radius of 0.85 Å, which is comparable to that of Ir$^{4+}$ (0.765 Å)[31]. Indeed, there are three known polymorphs of Na$_2$PrO$_3$, two of which have honeycomb[32-34] and hyperhoneycomb[35] lattices of edge-sharing Pr$^{4+}$O$_6$ octahedra as found in $\alpha$-[36] and $\beta$-[13, 14] Li$_2$IrO$_3$. In the latter the magnetic ions form a hyperhoneycomb lattice, which has the same local three-fold coordination as the planar honeycomb, but where additional bond rotations make it a three-dimensionally connected lattice. The third polymorph[37] of Na$_2$PrO$_3$ has a cubic rock salt structure with completely random arrangement of the cations in a 2:1 Na:Pr ratio on the same crystallographic site. Hyperhoneycomb $\beta$-Na$_2$PrO$_3$ was originally synthesized only in only a very small single crystal form and proposed to have a monoclinic space group of $C2/c$ with a full order of Na and Pr[35]. The structure of honeycomb $\alpha$-Na$_2$PrO$_3$ was proposed to be described by a (different) $C2/c$ unit cell with substantial disorder between Na and Pr because powder samples suffer from layer stacking faults and no single crystals have yet been reported[32, 34]. So far, physical properties have been investigated only in powder samples of $\alpha$-Na$_2$PrO$_3$[32, 34]. The synthesis and experimental investigation are both challenging because of the moisture sensitivity of Na$_2$PrO$_3$ and of the starting chemicals.

We have recently succeeded[38] in synthesizing $\beta$-Na$_2$PrO$_3$ by removing low-melting impurities that favor $\alpha$-Na$_2$PrO$_3$. In this study, we extend this strategy to obtain single crystals of $\alpha$-Na$_2$PrO$_3$ and we revise the crystal structure based on X-ray diffraction from a disorder-free, untwinned single crystal. We describe the synthetic protocols for separately obtaining all three polymorphs of Na$_2$PrO$_3$, provide a comprehensive study of the influence of synthesis parameters on the growth of single crystals of both α- and β-phases and compare the structural and magnetic similarities of both phases.

## EXPERIMENTAL SECTION

**Powder Synthesis of α- and β-Na$_2$PrO$_3$.** All samples were handled inside a nitrogen filled glovebox. Polycrystalline samples of α- and β-Na$_2$PrO$_3$ were synthesized either by annealing cubic Na$_2$PrO$_3$ or by a direct solid-state reaction between Na$_2$O$_2$ and Pr$_6$O$_{11}$. The cubic polymorph was first synthesized by a conventional solid-state reaction of Na$_2$O$_2$ (Alfa Aesar, 95%) and Pr$_6$O$_{11}$ (Merck Life Science, 99.9%). Because as-received Pr$_6$O$_{11}$ contains PrO$_2$ and unidentified phases, it was calcined in air at 800°C for 24 h to obtain phase-pure Pr$_6$O$_{11}$. In a typical synthesis, 1.3 mmol of Pr$_6$O$_{11}$ and 8.6 mmol of Na$_2$O$_2$, which amounts to a 10 mol% excess use of Na$_2$O$_2$, were thoroughly ground and pelletized. The pellet was loaded in an evacuated ($P < 1$ Pa) 40 cm long $\varphi$ = 17 mm diameter fused silica tube. The ampule was placed in a horizontal furnace and reacted at 400°C for 48 h. The heated sample contained purely the cubic phase, weighing 1.8 g. The heating time was determined such that no Na$_2$O$_2$ remained in the mixture after the reaction. The polycrystalline cubic phase sample was thoroughly ground, pelletized, and loaded in an open silver tube inside the glovebox. The silver tube was sealed inside an evacuated fused silica tube and reacted at 600 and 800°C for 12 h to obtain powder α- and β- phases, respectively. The direct solid-state reaction between Na$_2$O$_2$ and Pr$_6$O$_{11}$ was performed for both Na-deficient and Na-excess mixtures; the thoroughly ground powder was pressed into a pellet, loaded in an open silver tube, sealed in an evacuated fused silica tube, and reacted at higher temperatures. Samples obtained by annealing the cubic phase were used to measure the physical properties.

**Single Crystal Growth of α- and β-Na$_2$PrO$_3$.** All samples were handled inside a nitrogen filled glovebox except Pr$_6$O$_{11}$. Single crystals of α- and β-Na$_2$PrO$_3$ were synthesized by a solid-state reaction of Li$_2$O (Alfa Aesar, 99.5%), Na$_2$O (Alfa Aesar, 80%), and PrO$_2$. PrO$_2$ was synthesized by heating calcined Pr$_6$O$_{11}$ in an O$_2$ flow at 280°C for 4 weeks. Li$_8$PrO$_6$ was synthesized by heating stoichiometric mixture of as-received Pr$_6$O$_{11}$ and Li$_2$O in an O$_2$ flow at 700°C for 24 h. In a typical synthesis, 0.8 mmol of PrO$_2$, 1.6 mmol of Na$_2$O, and 3.2 mmol of Li$_2$O were ground and pressed into a $\varphi$ = 5 mm diameter pellet. The pellet was placed in a 5 cm long $\varphi$ = 6 mm diameter silver tube. One side of the silver tube was sealed with flame before the pellet was loaded and the other side was mechanically sealed inside the glovebox after the sample was loaded and finally securely sealed with flame. The pellet was sealed in an evacuated quartz tube. The ampule was placed in a box furnace and reacted at 750°C for more than 2 weeks and less than a month. Rhombic dark brown crystals of β-Na$_2$PrO$_3$ with a developed *ab* plane of the largest area of 0.5 mm$^2$ grew on the surface of the pellet and smaller crystals were found inside. The crystals were surrounded by yellow crystals of Li$_8$PrO$_6$ and colorless crystals of Na$_2$O. The rhombic crystals were mechanically extracted from the pellet. A small amount of α-Na$_2$PrO$_3$ formed as very thin crystals with a diameter up to 200 μm. Addition of NaOH made by hydration of Na$_2$O yielded only α-Na$_2$PrO$_3$ crystals.

**Characterization.** Powder X-ray diffraction (XRD) analysis was performed to confirm the purity of the samples using a Panalytical X'pert diffractometer (monochromatic Cu $K\alpha_1$, $\lambda$ = 1.54056 Å). To prevent contact with air the specimen was prepared in a nitrogen filled glovebox and placed in a dome shaped sample holder sealed inside the glovebox. The data were collected within a $2\theta$ range of 10 to 60° with a step of 0.02°. The FullProf Suite was used for structural refinement[39]. Single crystal XRD measurements were performed on an Oxford Diffraction Supernova diffractometer at room temperature with Mo $K\alpha$ radiation ($\lambda$ = 0.71073 Å). Crystals of the β-phase were covered with Dow Corning vacuum grease and no obvious decomposition was observed during the maximum measurement time of 2 h. Crystals of the α-phase were sealed under vacuum inside a glass capillary, after they had been coated with a thin layer of melted paraffin such that they could be stuck in a fixed position on the glass capillary. The X-ray data were collected and reduced using the CrysAlisPro software. Absorption corrections were made empirically. Initial models of the crystal structures were obtained with SIR-2014[40] and refined using Olex2[41]. Extinction corrections were applied by fitting the observed structure factor $|F_{obs}|$ to $|F_{calc}|/[1+\xi*0.001|F_{calc}|^2\lambda^3/\sin(2\theta)]^{1/4}$, where $\xi$ was refined. The atomic displacement parameters were refined anisotropically. VESTA was used for structure visualization[42].

**Magnetometry.** Magnetization measurements were performed using a Quantum Design MPMS3 system in fields up to 7 T and temperatures down to 2 K, first on powders and subsequently on co-aligned single crystals of the β-phase with a total combined mass of ~1 mg on a diamagnetic substrate. The single crystal susceptibilities showed contributions from a ferromagnetic impurity phase Pr$_6$O$_{11}$ due to decomposition on the surface, which was subtracted by comparing the magnetization data to single-crystal torque magnetometry data, which will be reported elsewhere.

## RESULTS AND DISCUSSION

**Crystal Growth.** Tables I and II summarize various synthetic conditions for obtaining powder and single crystals of α- and β-Na$_2$PrO$_3$, respectively. We first describe the powder synthesis in detail and then discuss single crystal growth. In previous studies of Na$_2$PrO$_3$, the powder sample of α-Na$_2$PrO$_3$ was prepared by the reaction of Pr$_6$O$_{11}$ with Na$_2$O$_2$ or Na$_2$O under an O$_2$ flow condition[32-34] at 600°C-700°C. Two reaction routes[35] under sealed conditions were employed for β-Na$_2$PrO$_3$, namely an ion exchange reaction between Na$_2$O and Li$_8$PrO$_6$ at 700°C or a direct solid-state reaction of Pr$_6$O$_{11}$ with a slight excess of Na$_2$O$_2$ at 650°C. The critical difference between the O$_2$ flow reaction and the sealed condition is contamination of Na$_2$O or Na$_2$O$_2$ by water. A continuous flow of oxygen is expected to introduce substantial amounts of moisture over time and to produce NaOH. To prevent the degradation of sodium oxides, we used sealed conditions except for the purpose of reproducing the previous studies. The basic trend found in this study was that the melt stabilizes α-Na$_2$PrO$_3$ although β-Na$_2$PrO$_3$ seems to be the more stable phase, which can be understood from Ostwald ripening of crystal growth. NaOH and Na$_2$O$_2$ melt at 318 and 675°C, respectively, meaning that they need to be consumed below the temperature at which the β-phase starts to form (~550°C). Although the direct reaction of Na$_2$O$_2$ and Pr$_6$O$_{11}$ seems to be simplest, it results in a mixture of α- and β-Na$_2$PrO$_3$ for the above-mentioned reason. The most reproducible and scalable procedure for producing the β-phase is to make the pure cubic phase with low crystallinity at a low temperature and anneal it at 800°C; a lower temperature of 600°C resulted in the α-phase. The powder diffraction pattern of each phase is presented in Figure 1.

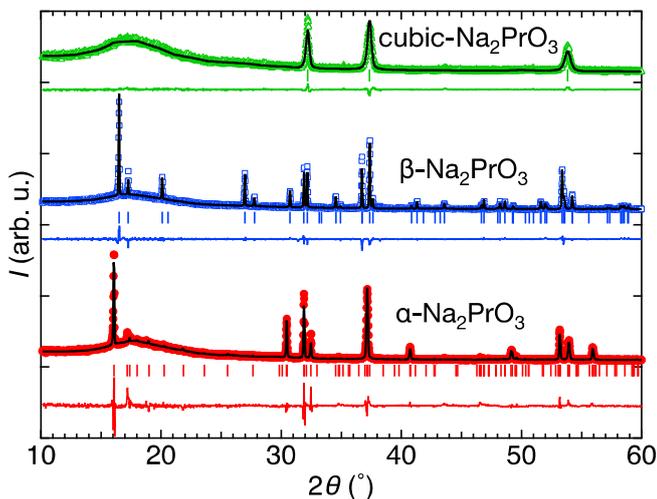

**Figure 1**. Powder XRD patterns of the three polymorphs of Na$_2$PrO$_3$, offset vertically for the sake of clarity (data reproduced from ref.[38]). Red circles, blue squares, and green triangles represent the powder diffraction patterns of α-, β-, and cubic-Na$_2$PrO$_3$, respectively. Black lines and bars under each data set indicate fits to Rietveld refinement and positions of Bragg peaks of each phase. Red, blue, and green lines are the residuals of the fits of each phase. Details of the refinements are presented in Table III with structural parameters listed in Tables IV and V.

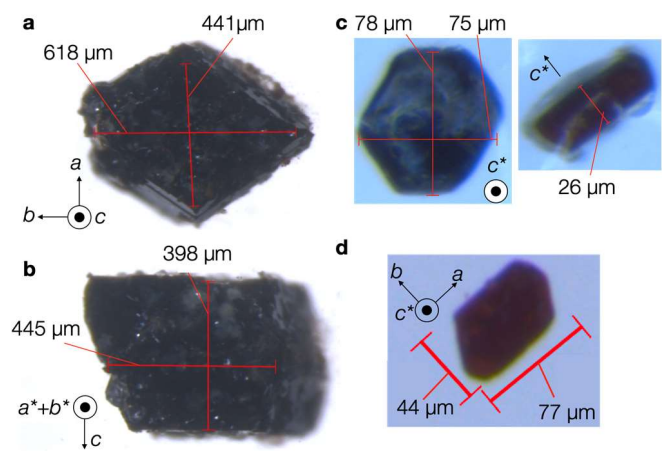

**Figure 2**. Optical microscope images of single crystals of α- and β-Na$_2$PrO$_3$. **a**, Largest grown β-Na$_2$PrO$_3$ crystal viewed along the (001) axis. **b**, Same β-Na$_2$PrO$_3$ crystal viewed along the (110) axis. **c**, Hexagonally shaped twinned crystal of α-Na$_2$PrO$_3$ viewed along the $c^*$ axis (left image) and a perpendicular direction (right image). The color appears denser when viewed perpendicular to the honeycomb plane. **d**, Single grain, untwinned α-Na$_2$PrO$_3$ crystal with an elongated shape along the monoclinic $a$ axis.

For single crystal growth of α- and β-Na$_2$PrO$_3$, all of the attempts to crystallize them by a flux method failed; a mixture of Pr$_6$O$_{11}$ and excess NaF and Na$_2$O$_2$ formed crystals of the cubic phase. As noted in the previous study[35], the intermediate compound Li$_8$PrO$_6$ is key to stabilization of the β-phase. More specifically, Li$_2$O works as a mineralizer in the formation of the β-phase because it was found that addition of Li$_2$O converts the α-phase to the β-phase. The largest single crystal of β-Na$_2$PrO$_3$ was prepared by the reaction of Li$_2$O, Na$_2$O, and PrO$_2$ instead of using Li$_8$PrO$_6$. The advantage of PrO$_2$ over Li$_8$PrO$_6$ is its insensitivity to moisture. Although Li$_8$PrO$_6$ slowly decomposes in moist air, it is typically prepared under oxygen flow conditions and inevitably contains a small amount of LiOH. Such a low melting hydroxide is unfavorable for formation of the β-phase. Single crystals of α-Na$_2$PrO$_3$ were found as a minority phase in the β-phase or predominantly formed by addition of NaOH. Figure 2 presents pictures of single crystals of each phase. Large crystals of the β-phase have a rhombic shape with each edge corresponding to one of the **a**±**b** unit cell diagonals (Figs. 2a-b). The α-phase crystals always form as thin platelets, hexagonal ones as shown in Fig. 2c (left image) tend to contain multiple structural twins rotated by 120° around a common $c^*$ axis. Single grain, untwinned α-crystals are elongated along the monoclinic $a$ axis with a typical crystal shape morphology shown in Fig. 2d. Small crystals of both phases look similar but only the α−phase crystals display significant pleochroism, they appear to be transparent brown when viewed along an in-plane direction (⊥$c^*$) but very dark in the out-of-plane direction (//$c^*$).

**Table I**. Selected synthetic conditions for obtaining polycrystalline powder of the three polymorphs of $Na_2PrO_3$, bold highlighting indicates the conditions used to produce the highest-purity $β-Na_2PrO_3$ powder used for the x-ray diffraction in Fig. 1 (bottom trace) and magnetic susceptibility in Fig. 8. Evacuated quartz tubes are used for reaction vessels in addition to open silver tubes inside the quartz tubes for the reaction temperature higher than 500°C.

| starting materials | reaction conditions | phase analysis |
|---|---|---|
| $xNa_2O_2$ ($x > 1$), $1Pr_6O_{11}$ | 350-450°C, 24-96 h | Cubic-$Na_2PrO_3$, $Na_2O_2$ |
| $xNa_2O_2$ ($x > 1$), $1Pr_6O_{11}$ | 550-640°C, 12 h | $α$- and $β$-$Na_2PrO_3$, NaOH, $PrO_2$ |
| $xNa_2O_2$ ($x < 1$), $1Pr_6O_{11}$ | 800°C, 12 h | $β$-$Na_2PrO_3$, NaOH, $PrO_2$ |
| $xNa_2O_2$ ($x > 0.1$), 1cubic-$Na_2PrO_3$ | 550°C-800°C, 12 h | $α$-$Na_2PrO_3$ |
| $xNa_2O$ ($x > 0.1$), 1cubic-$Na_2PrO_3$ | 550°C-800°C, 12 h | $α$-$Na_2PrO_3$ |
| **$xNa_2O_2$ ($x < 0.1$), 1cubic-$Na_2PrO_3$** | **800°C, 12 h** | **$β$-$Na_2PrO_3$, NaOH** |
| $0.1Li_2O$, $α$-$Na_2PrO_3$ | 800°C, 12 h | $β$-$Na_2PrO_3$, $Li_2O$ |

**Table II**. Selected synthetic conditions for obtaining single crystals of the three polymorphs of $Na_2PrO_3$, bold highlighting indicates the conditions used to obtain the largest size $β$-$Na_2PrO_3$ crystal shown in panels a and b of Figure 2.

| starting materials | reaction conditions | single crystals |
|---|---|---|
| $xNa_2O$ ($x < 2$), $1Li_8PrO_6$ | 650-750°C, 2-8 weeks, sealed Ag | $α$- and $β$-$Na_2PrO_3$ |
| $xNa_2O$ ($x > 2$), $1Li_8PrO_6$ | 650-750°C, 2-8 weeks, sealed Ag | unknown phase ($Na_xPr_yO$ with $x/y>2$) |
| $1Na_2O_2$, $1Li_8PrO_6$ | 650°C, 2-8 weeks, sealed Ag | $α$- and $β$-$Na_2PrO_3$ |
| $1Na_2O_2$, $1Li_8PrO_6$ | 700°C, 2-8 weeks, sealed Ag | $α$-$Na_2PrO_3$ |
| $2Na_2O_2$, 2NaF, $1PrO_2$ | Cooling from 700°C by -5°C/hr, sealed Ag | cubic-$Na_2PrO_3$ |
| **$xNa_2O$ ($x < 2$), $4Li_2O$, $1PrO_2$** | **650-750°C, 2-8 weeks, sealed Ag** | **$α$- and $β$-$Na_2PrO_3$** |
| $1.1Na_2O_2$, $1Li_8PrO_6$ | 650-750°C, 2-8 weeks, $O_2$ flow on Au foil | $α$- and $β$-$Na_2PrO_3$, $Li_2PrO_3$ |

**Table III**. Rietveld refinement results for the powder X-ray data on the three phases of $Na_2PrO_3$ plotted in Fig. 1. The atomic positions and isotropic $U_{iso}$ values are fixed to those values obtained from the single crystal refinement results given in Tables IV and VI. Calculations of the pattern of the $α$-phase was performed for the partially disordered structure in which the Na1 and Pr1/Pr2 sites are partially occupied by Pr and Na, respectively, as per Table V, with the positions of the other atoms fixed to those in Table IV. The refined partial occupancy of Na1 and Pr1/Pr2 sites is 54.7(6)% and 77.4 (3)% respectively[32]. In the cubic phase, Na/Pr and O are placed in the 4$a$ (0 0 0) and 4$b$ (1/2 1/2 1/2) positions, respectively, in which the occupancy of 4$a$ site by Na and Pr is fixed to 67% and 33%, respectively. For $β$-$Na_2PrO_3$ we used the structural model in Table VI. Use of the dome-shaped sample holder caused a substantial increase in the background signal, leading to higher $R_{exp}$ after background correction. $R_{exp}$ without background correction is 8.14%, 5.78, and 8.49 for $β$, $α$, and cubic phases, respectively.

| phase | $R_p$, $R_{wp}$, $R_{exp}$ (%) | space group and unit cell parameters |
|---|---|---|
| $α$-$Na_2PrO_3$ | 34.9, 27.3, 15.4 | $C2/c$, $a$ = 5.9586(2) Å, $b$ = 10.3297(3), $c$ = 11.2063(3) Å, $β$ = 100.328(1)° |
| $β$-$Na_2PrO_3$ | 27.5, 24.3, 16.6 | $Fddd$, $a$ = 6.7613(3) Å, $b$ = 9.7812(4) Å, $c$ = 20.5313(9) Å |
| cubic-$Na_2PrO_3$ | 22.5, 22.3, 15.0 | $Fm\bar{3}m$, $a$ = 4.8166(6) Å |

**Crystal Structure Determination.** The crystal structure of $α$-$Na_2PrO_3$ was determined by using single crystal X-ray diffraction data at room temperature. The results of the structural refinement are summarized in Table IV. We first describe the revised $α$-phase crystal structure and then compare the local coordination in both $α$- and $β$-structures.

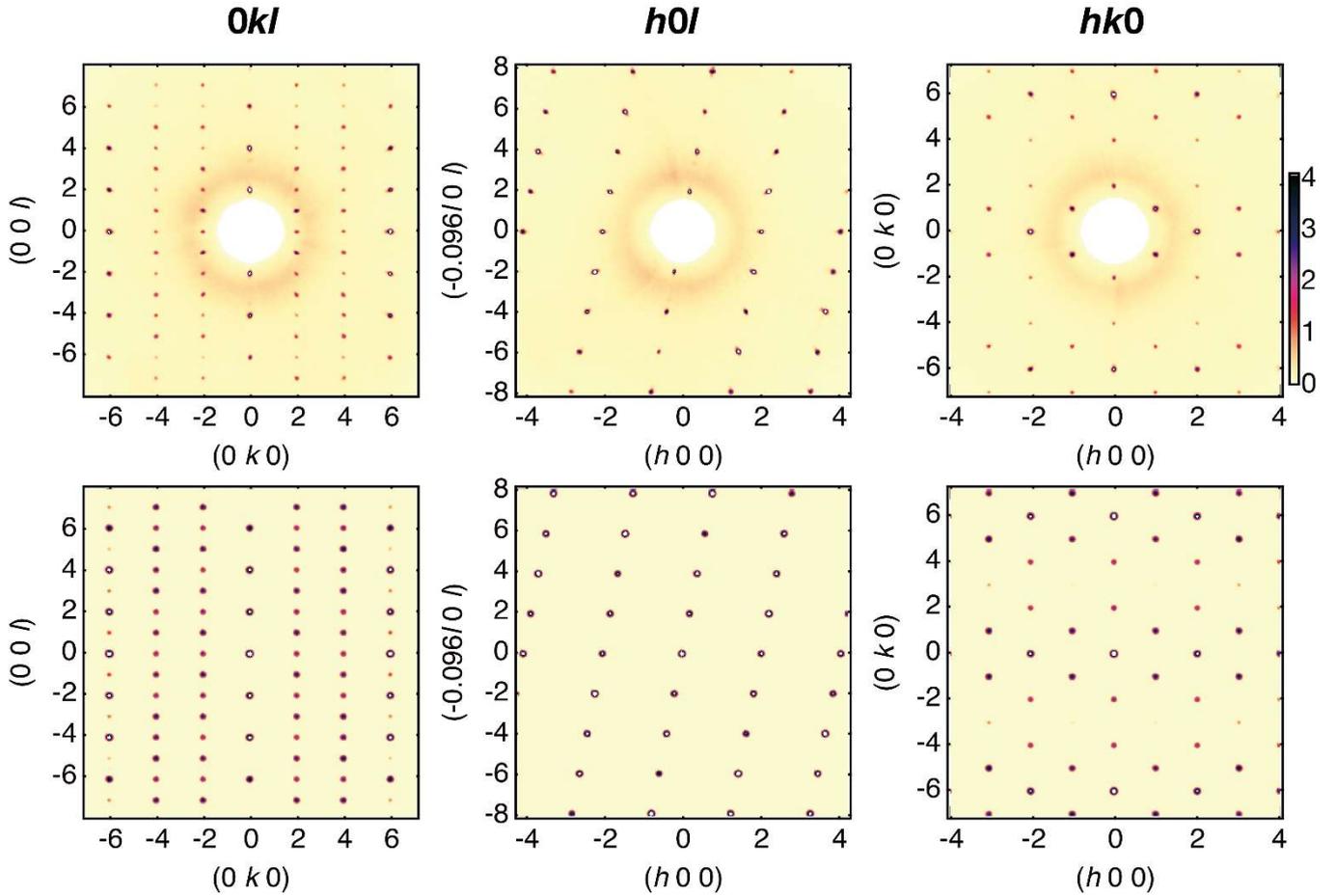

**Figure 3.** X-ray diffraction patterns of an untwinned α-Na$_2$PrO$_3$ single crystal. (top) Observed patterns in the 0*kl*, *h*0*l*, and *hk*0 planes in the region with scattering wavevector |Q| < 4.5 Å$^{-1}$. The plotted colour is $log_2(1 + I/255)$, where *I* is the observed diffraction intensity. (bottom) Corresponding calculated X-ray diffraction patterns based on the refined structural model in Table IV. *C*-centering requires the sum $h + k$ to be even and the *c*-glide (mirror in the (*x*0*z*) plane followed by translation by **c**/2) forbids odd *l* reflections in the *h*0*l* plane. The halo around the origin is due to scattering from the vacuum sealed glass capillary inside which the crystal was placed. All observed diffraction peaks are sharp with no detectable diffuse scattering rods along *l*, indicating the absence of layer stacking faults, common in powder samples of α-Na$_2$PrO$_3$[32,34] as well as other layered honeycomb systems such as Na$_2$IrO$_3$[43].

For the α-phase, elongated crystals as in Figure 2d showed no apparent twinning or stacking faults. The previous powder study[32,34] proposed a monoclinic space group of *C*2/*c* with the following unit cell lattice parameters $a_p$ ~ 5.96 Å, $b_p$ ~ 10.3 Å, $c_p$ ~ 11.7 Å, and $β_p$ ~ 110°. In our study, we use a revised monoclinic unit cell with an angle *β* of ~100°, which is much closer to 90°. Revised unit cell basis vectors ***a, b, c*** are related to previously proposed[32,34] unit cell basis vectors $a_p$, $b_p$, $c_p$ as follows: ***a*** = $a_p$, ***b*** = −$b_p$, and ***c*** = −$a_p$ − $c_p$. This transformation does not involve a space group change. To illustrate the validity of the revised unit cell, single crystal X-ray diffraction patterns in the 0*kl*, *h*0*l*, and *hk*0 planes are presented in Fig. 3 (top row). Throughout we index reflections in terms of Miller indices with reference to the crystallographic (revised) unit cell. The diffraction patterns have 2-fold rotational symmetry around (010) (normal to panels in the middle column of Figure 3) and mirror symmetry normal to this axis (see the diffraction patterns in the left and right columns), leading to a 2/*m* Laue class. All observed peaks satisfy the $h+k$ = even selection rule, as expected for a *C*-centered cell. These two considerations narrow down the possible space groups to *C*2/*c* and *C*2/*m*. Those could be differentiated by the pattern in the *h*0*l* plane (Fig 3. middle column), which clearly reveals systematic absences when *l* is odd, as expected for a *C*2/*c* space group and attributed to the *c*-glide. The correct choice for the *c* lattice parameter is demonstrated by the presence of odd *l* reflections in the 0*kl* diffraction pattern (left column).

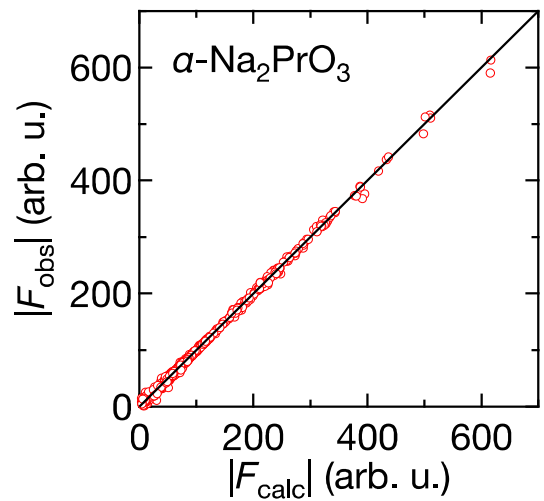

**Figure 4.** Observed vs calculated X-ray structure factor for the refined structural model for α-Na$_2$PrO$_3$ in Table IV. Solid line shows the 1:1 agreement.

Crystal structure refinement of the α-phase converged to the expected Pr honeycomb structure in the monoclinic *ab* plane

(see Fig. 5a) and the obtained atomic coordinates are listed in Table IV. This revised crystal structure of α-Na$_2$PrO$_3$ is isostructural to Na$_2$TbO$_3$[35]. The calculated diffraction patterns agree well with the observed patterns as Figure 3 shows, upon comparison of the top and bottom rows. Observed and calculated X-ray structure factors are quantitatively compared in Fig. 4. The refined crystal structures of both α- and β-Na$_2$PrO$_3$ are schematically illustrated in Figure 5. Both polymorphs have ordered rock salt structures with no disorder in the arrangement of sodium and praseodymium. On the basis of our single crystal results in α-Na$_2$PrO$_3$, two-dimensional honeycombs form by edge-sharing of two structurally non-equivalent PrO$_6$ (dark/light) blue octahedra illustrated in Fig. 5a. Compared with honeycomb iridate α-Li$_2$IrO$_3$ with a single honeycomb layer per unit cell and single Ir site, the crystal structure of α-Na$_2$PrO$_3$ has a two layer stacking periodicity, with the two layers in the unit cell related by a c-glide (mirror in the ac plane followed by translation by c/2), so the interlayer spacing is roughly doubled compared to that of α-Li$_2$IrO$_3$. One of the sodium sites Na1 sits at the center of the honeycombs and the Na2 and Na3 sites form a hexagonal lattice in-between stacked honeycomb layers as plotted in Fig. 5b. The displacement parameters of sodium atoms are much more anisotropic than those of the other atoms; the thermal ellipsoid of Na1 is elongated toward c* and those of Na2 and Na3 are more elongated in-plane presumably because Na is less likely to move toward the heavy Pr atoms.

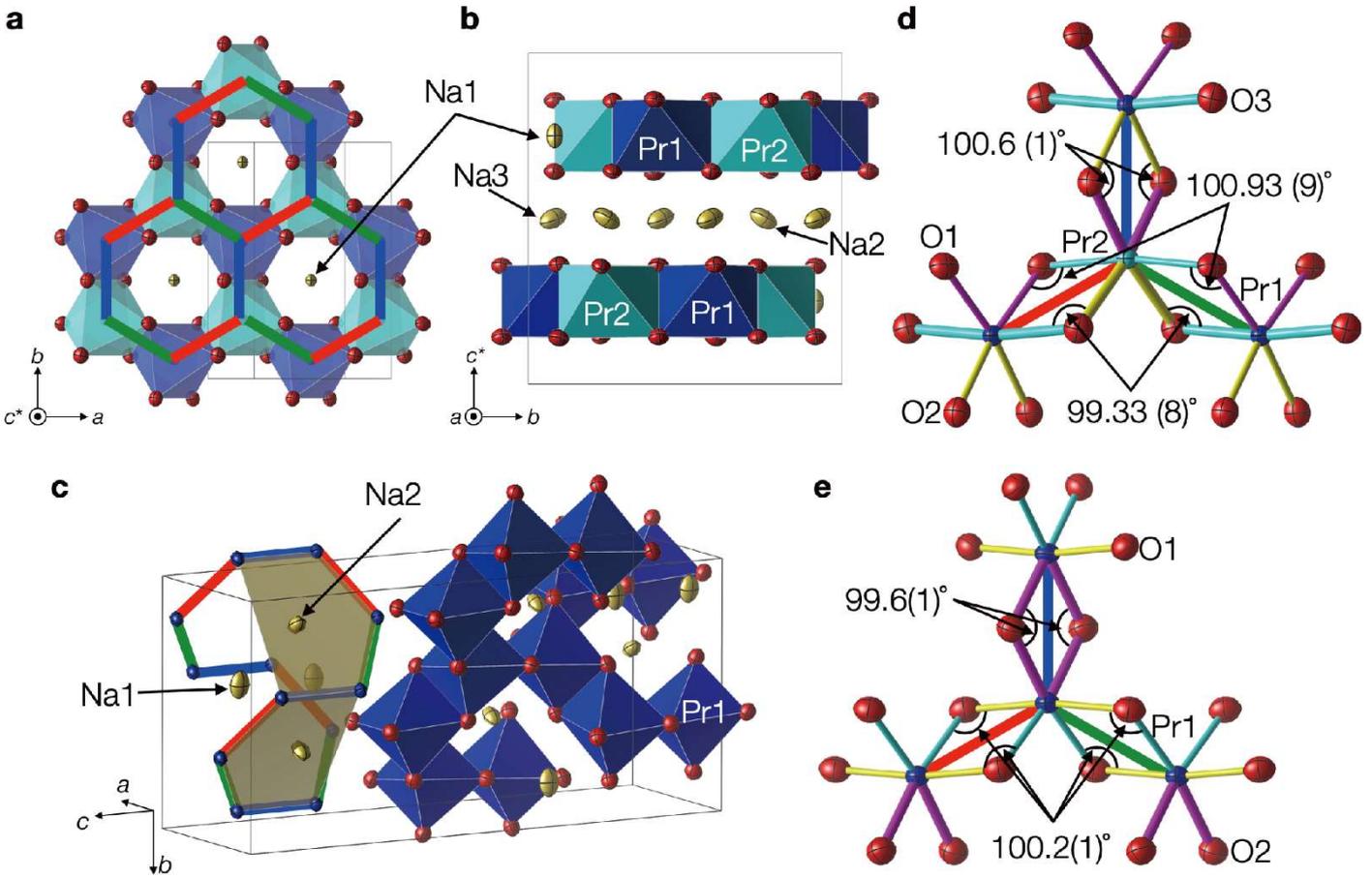

**Figure 5**. Crystal structures of α- and β-Na$_2$PrO$_3$. Light and dark blue, yellow and red ellipsoids represent praseodymium, sodium, and oxygen atoms, respectively, scaled by their anisotropic displacement parameters $U_{ij}$. The probability that nuclei are included in the interior of the ellipsoid is 99%. (a) Connectivity of PrO$_6$ in the α-phase in the ab plane. Alternating Pr1 and Pr2 sites form a honeycomb network. The unit cell outline is shown by thin black lines. (b) Stacking of the honeycomb planes in the α-phase. (c) Connectivity of PrO$_6$ octahedra in the β-phase (using atomic positions from ref.[38]). Red, green and blue bond coloring in all panels correspond to the bond Ising **x**, **y** and **z** character in a Kitaev model for an ideal structural, in which all octahedra are cubic and the three Pr-O2-Pr superexchange planes meeting at a site are mutually orthogonal, with the **x**, **y** and **z** axes normal to those planes. (d) Edge-sharing connectivity of PrO$_6$ octahedra in d) the α-phase and e) β-phase with bond angles indicated. Inequivalent Pr-O bonds are shown in different thickness and color. In panels d) and e) two of the Pr-Pr bonds that share a site are symmetry-related (red and green) by a 2-fold axis along the third Pr-Pr bond (blue) that shares the same site.

For the sake of completeness we note than an ideal version of the crystal structure of α-Na$_2$PrO$_3$ with all octahedra cubic is obtained by a slight rescaling of the lattice parameters to be in the ratio $a:b:c = \sqrt{6}:3\sqrt{2}:\sqrt{22}$, angle $\beta = \cos^{-1}(-1/\sqrt{33}) = 100.0°$ and ideal atomic fractional coordinates Na1(0, 11/12, 1/4), Na2(1/4, 1/4, 0), Na3(1/4, 7/12, 0), Pr1(0, 7/12, 1/4), Pr2(0, 1/4, 1/4), O1(1/8, 3/4, 1/8), O2(1/8, 5/12, 1/8), and O3(1/8, 1/12, 1/8) with the set of axes **x** = $-3\sqrt{2}\mathbf{a}/8b + \sqrt{2}\mathbf{b}/2b + 3\sqrt{2}\mathbf{c}/8b$, **y** $= -3\sqrt{2}\mathbf{a}/8b - \sqrt{2}\mathbf{b}/2b + 3\sqrt{2}\mathbf{c}/8b$ and **z** $= 9\sqrt{2}\mathbf{a}/8b + 3\sqrt{2}\mathbf{c}/8b$, normal to the three Pr-O-Pr superexchange planes meeting at every site.

In previous reports, Rietveld refinement of the α phase structure on powder sample proposed that 1/3 of one of the two Pr sites in the honeycomb plane was occupied by Na and the other Pr site was fully occupied by Pr[32], while local ordering of Na and Pr was suggested by pair distribution function (PDF) analysis[34]. This local structure suggested by the PDF analysis[34] extended to the case of no atomic site mixing (i.e. both nominally Pr sites fully occupied by Pr) is different from the structure in Table IV in that the positions of Na1 and Pr1 are swapped, however such a structure with those two sites swapped is not compatible with our single crystal X-ray data. Attempts to refine our data in this structure increased $R_{wp,all}$ from 3.87% to 4.33% and resulted in unphysical, nonpositive definite anisotropic displacement parameters.

With regard to the powder diffraction patterns, our data show the same disorder effects as previous studies. Many diffraction peaks expected from the fully ordered $C2/c$ structure found in single crystals were not observed in the powder XRD data as shown in Fig. 6 (the dashed line is the expected pattern using the fully-ordered crystal structure in Table IV). Indication of the $C2/c$ structure appears only around the (1 1 0) position with $d$~5.1 Å, which is forbidden in a single-layer $C2/m$ model, isostructural to $\alpha$-Li$_2$IrO$_3$. Unit cell basis vectors of this single layer $C2/m$ structure; $a_s$, $b_s$, and $c_s$, are related to those of the two-layer $C2/c$ structure by the following relations $a_s = a$, $b_s = –b$, and $c_s = –a/2 – c/2$. The powder pattern presented in Figs. 1 and 6 was fitted by allowing some finite occupation of Na on both of the two Pr sites and a finite occupation of Pr on the Na1 site in the honeycomb layer of the crystal structure of $\alpha$-Na$_2$PrO$_3$ to mimic the effect of layer stacking faults. The site-mixing model described above leads to the suppression or disappearance of many diffraction peaks and a significant decrease of the occupancy of the honeycomb center Na1 site by almost half, as summarized in Table V.

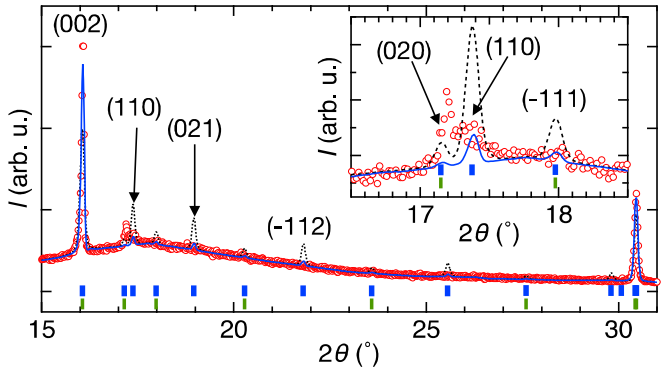

**Figure 6.** Diffraction pattern of the polycrystalline $\alpha$-Na$_2$PrO$_3$ phase shown by red points, compared with patterns calculated for the fully ordered structure in Table IV (black dotted line) and a partially ordered structure (blue line, same as in Fig. 1) as described in the text. Positions of Bragg peaks of the two-layer $C2/c$ and single-layer $C2/m$ structures described in the text are shown below the pattern in thick blue and thin green bars, respectively. The inset shows a close-up of the region near the position of the (110) diffraction peak disallowed in the $C2/m$ structure. The peak indexing is with reference to the two-layer $C2/c$ unit cell.

With regard to the crystal structure of the $\beta$-phase, a monoclinic unit cell with space group $C2/c$ and lattice parameters $a_m$ ~ 6.79 Å, $b_m$ ~ 9.77 Å, $c_m$ ~ 10.8 Å, and $\beta$ ~ 108° was originally proposed in the original synthesis study[35]. We recently revised the crystal structure model on the basis of analysis of an extensive single crystal x-ray diffraction data set. We find an orthorhombic unit cell of space group $Fddd$ (a supergroup of $C2/c$) with the following lattice parameters $a$ ~ 6.79 Å, $b$ ~ 9.77 Å, and $c$ ~ 20.6 Å. This revised crystal structure is isostructural to $\beta$-Li$_2$IrO$_3$, with refinement results presented in Table VI. $\beta$-Na$_2$PrO$_3$ features a full three-dimensionally linked network of PrO$_6$ octahedra connected via edge-sharing to form a hyperhoneycomb lattice. There is only one crystallographic Pr site and the hyperhoneycomb lattice comprises two families of zigzag chains running alternatingly along the basal plane diagonal $a + b$ and $a – b$ directions, connected by bonds along the $c$ axis. Theoretical models that discuss the possible realization of a Kitaev spin model on the hyperhoneycomb lattice are based on an ideal crystal structure with all PrO$_6$ octahedra regular (cubic) and all Pr-O-Pr angles equal to 90°, which can be obtained from the actual crystal structure in Table VI by scaling slightly the lattice parameters to have ratios $a:b:c$ =1:$\sqrt{2}$:3 and displacing the ions slightly from their actual to ideal positions (for details see ref.[38]). For this ideal structure the three Pr-O$_2$-Pr superexchange planes that meet at a Pr site are mutually orthogonal with normals along a Cartesian set of x,y and z axes expressed as **x** = (–*a/a* - *c/c*)/$\sqrt{2}$, **y** = (*a/a* - *c/c*)/$\sqrt{2}$ and **z**= -*b/b*, in relation to the idealized unit cell basis vectors. In the Kitaev model, each bond carries an Ising interaction between the magnetic moment components normal to the Pr-O$_2$-Pr plane of the bond. One can then color code all the bonds red/green/blue according to the **xyz** Ising character and we illustrate this by the color coding of the Pr-Pr bonds in panels c and e of Figure 5. Two crystallographically inequivalent sodium sites Na1 and Na2 fill the hollow of the plaquette. Na2 is located inside the incomplete honeycomb sections (shaded polygons in Fig 5c) formed by adjacent **z** -bonds connected via zigzag bonds and its displacement features out-of-plane motion as in the $\alpha$-phase; Na1, away from the incomplete honeycombs, has an ellipsoid elongated mainly along the $b$ axis.

Despite the very different three-dimensional networks formed by the Pr ions in the two polymorphs, the local structural environments around the Pr ions closely resemble each other. Table VII presents the Pr-O bond lengths and Pr-O-Pr bond angles for $\alpha$- and $\beta$-Na$_2$PrO$_3$. Figs. 5d and e illustrate the local threefold-coordinated geometry in the two phases. The Pr-Pr distances show relatively small variation in the range 3.422-3.444 Å, which is close to the corresponding value in the disordered cubic structure $a_c/\sqrt{2}$ ~ 3.406 Å. The Pr-O-Pr bond angles also take similar values for all the bonds, in the range 99-101°. The normals to the Pr-O$_2$-Pr planes that share a site make relative angles of 92.5° between **x-y** bonds and 91.8° between **y-z** and **z-x** bonds in the $\beta$ phase and 91.5-92.1° for **x-y** bonds and 91.7°-92.0° between **y-z** and **z-x** bonds in the $\alpha$ phase. The slight deviations of all of those angles away from the ideal values of 90° indicate slight departures of the actual crystal structures from the ideal structures with cubic octahedra, where the three bonds sharing a site are connected by a local (pseudo) three-fold axis normal to the plane of the three bonds.

As the deviation of the bond angles from the ideal value of 90° is expected to lead to suppressed Kitaev interactions, Fig. 7 compares the degree of distortion in candidate Kitaev honeycomb and hyperhoneycomb magnets. The deviation of bond angles in Na$_2$PrO$_3$ is comparable to Na$_2$IrO$_3$ and significantly larger than that in α- and β-Li$_2$IrO$_3$ or α-RuCl$_3$. The distortion roughly correlates with the difference in the ionic radii of the two cations in the rock salt structure; the ionic radii of the constituting elements are 0.76 Å, 0.765 Å, 0.85 Å, and 1.02 Å, for Li$^+$, Ir$^{4+}$, Pr$^{4+}$, and Na$^+$, respectively, in octahedral (6-fold) coordination.

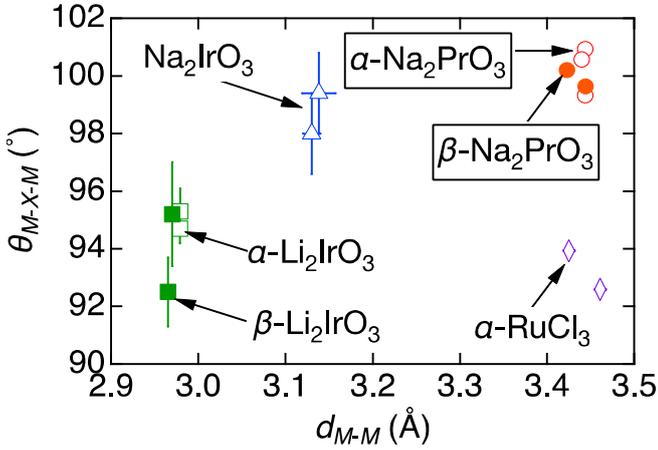

**Figure 7.** Comparison of bond lengths and angles of candidate Kitaev honeycomb and hyperhoneycomb materials. The distance between the nearest neighbor magnetic ions $d_{M-M}$ and the bond angle $\theta_{M-X-M}$ between two metal ions $M$ and the coordinated anion $X$ are plotted for $M$ = Pr$^{4+}$, Ir$^{4+}$, Ru$^{3+}$, and $X$ = O$^{2-}$, Cl$^-$. Green squares, blue triangles, and purple diamonds represent data for Na$_2$PrO$_3$, α- and β-Li$_2$IrO$_3$[14, 44], Na$_2$IrO$_3$[43], and α-RuCl$_3$[45] and empty and filled red circles correspond to α- and β- Na$_2$PrO$_3$, respectively. Only the values of α-RuCl$_3$ are measured at 80 K, all others are taken from room temperature data.

**Magnetic Properties of α- and β-Na$_2$PrO$_3$**

Powder and single crystal magnetization measurements were performed to investigate the magnetic properties of the β-phase. Susceptibility follows a Curie-Weiss law above 20 K. The effective moment and Weiss temperatures are 0.80(1)μ$_B$ and -15(1) K, respectively. For the α-phase, a Curie-Weiss fit above 50 K yielded the effective moment and Weiss temperature of 1.08(1) μ$_B$ and –16(1) K, respectively, which are similar to the values obtained in the previous study[32]. The observed effective moment is reduced from the value of $\left(\frac{10}{7}\right) * \sqrt{3} / 2$ μ$_B$ = 1.24 μ$_B$, expected for the pure Γ$_7$ ground state doublet in the limit of very weak octahedral cubic crystal field relative to the spin orbit coupling[46]. Furthermore the spin interaction is large as insulating rare-earth compounds typically have exchanges on the order of 1 K[19]. These data suggest the presence of a large hybridization between praseodymium and oxygen atoms because of the unusually high Pr valence. Recent studies found comparable energy scales of the crystal field and spin-orbit coupling in Pr$^{4+}$ oxides[47, 48], which leads to mixing of the nominally $j_{eff}$ = 1/2, Γ$_7$ doublet derived from the free-ion $J$ = 5/2 multiplet with higher energy states derived from the $J$ = 7/2 multiplet even in the cubic limit, with further mixing of more states when the local environment is distorted away from cubic, which is the case in both α- and β-Na$_2$PrO$_3$.

Temperature-dependent magnetic susceptibility measurements of β-phase powder and single crystals exhibit the same magnetic ordering temperature of 5.2 K as shown in Fig.8b. This excludes the possibility of contamination of Na sites with Li and evidences the good stoichiometry of our samples. Single crystal magnetization showed clear anisotropy below $T_N$ of 5.2 K, suggesting that spins are predominantly aligned along the $a$ axis. In agreement with previous reports for α-phase powder samples, our field-cooled susceptibility data exhibits a small anomaly at $T_N$ = 4.6 K, attributed to the onset of a weak ferromagnetic component[32,33]. More detailed investigations of the magnetic properties of the two phases will be reported in ref.[38] and elsewhere.

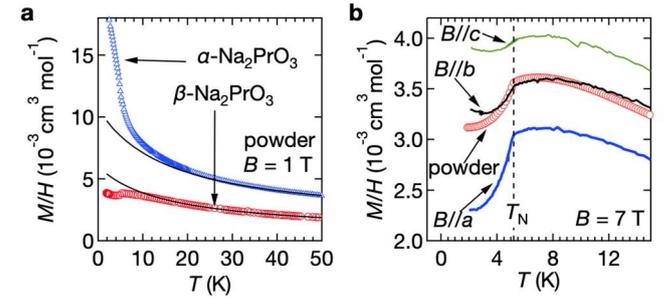

**Figure 8** (a) Temperature dependence of magnetic susceptibility of α- and β-Na$_2$PrO$_3$ measured by field cooling at 1 T. Black lines are fits to a Curie Weiss law. (b) Susceptibilities of powder and single crystals measured along $a$, $b$, and $c$ axes of the β-phase at 7 T.

## CONCLUSIONS

We have succeeded in the synthesis of powder and single crystals of candidate Kitaev magnets honeycomb α-Na$_2$PrO$_3$ and hyperhoneycomb β-Na$_2$PrO$_3$ by performing reactions under inert conditions and identification of the mineralizing effect of NaOH and Li$_2$O for the α- and β-polymorphs, respectively. Single crystals of both polymorphs show fully-ordered structures with no site mixing or structural stacking faults. Revised crystal structures are obtained for both polymorphs from refinement of single crystal X-ray diffraction data. In both structures Pr$^{4+}$ ions realize threefold-coordinated lattices of spin-orbit entangled magnetic moments inside edge sharing octahedra, as required for Kitaev and related models with anisotropic bond-dependent exchanges. The different character of the orbitals mediating the superexchange interactions as well as the different local octahedral distortions could potentially realize distinct regimes of frustrated bond-dependent spin Hamiltonians compared with the much-explored 4$d$ and 5$d$ Kitaev materials.

## ACKNOWLEDGEMENTS

This research was supported by the European Research Council under the European Union's Horizon 2020 Research and Innovation Programme Grant Agreement Number 788814 (EQFT) and the Engineering and Physical Sciences Research Council (EPSRC) under grant No. EP/M020517/1. RO acknowledges support from JSPS KAKENHI (Grant No. 23K19027JST) and JST ASPIRE (Grant No. JPMJAP2314).

## Data Availability

**Table IV**. Crystallographic data and the results of structural refinement for $\alpha$-Na$_2$PrO$_3$ from single-crystal X-ray data at 293(2) K. Peaks with $I > 2\sigma(I)$ are defined as observed ones. We use the $C2/c$ space group setting with unique axis $b$ and cell choice 1 (origin at $\bar{1}$ on $c$-glide plane). Atomic fractional coordinates, and equivalent isotropic $U_{iso}$ and anisotropic $U_{ij}$ displacement parameters (in units of $10^{-3}$Å$^2$) with estimated standard deviations in parentheses.

| Formula | $\alpha$-Na$_2$PrO$_3$ |
|---|---|
| Space group | $C2/c$ |
| $a$ (Å) | 5.9559(2) |
| $b$ (Å) | 10.3371(3) |
| $c$ (Å) | 11.2069(4) |
| $\beta$ (°) | 100.423(3) |
| $V$ (Å$^3$) | 678.59(3) |
| $Z$ | 8 |
| $\mu$ (mm$^{-1}$) | 14.418 |
| $F_{000}$ | 840 |
| $R_{p,obs}$ | 1.94 |
| $R_{p,all}$ | 2.31 |
| $R_{wp,obs}$ | 3.69 |
| $R_{wp,all}$ | 3.87 |
| $R_{int}$ | 4.29 |
| Index range | $-8 < h < 7$ |
| | $-14 < k < 13$ |
| | $-15 < l < 15$ |
| $N_{measured}$ | 8432 |
| $N_{unique}$ | 900 |
| $N_{obs}$ | 837 |
| $N_{param}$ | 59 |
| Crystal size ($\mu$m$^3$) | 100×50×10 |
| extinction coefficient $\xi$ | 6.9(2) × 10$^{-3}$ |

| Atom | Wyckoff | $x$ | $y$ | $z$ | $U_{iso}$ | $U_{11}$ | $U_{22}$ | $U_{33}$ | $U_{12}$ | $U_{13}$ | $U_{23}$ |
|---|---|---|---|---|---|---|---|---|---|---|---|
| Pr1 | 4$e$ | 0 | 0.58165(2) | 1/4 | 5.21(8) | 4.9(1) | 4.4(1) | 6.2(1) | 0 | 0.7(1) | 0 |
| Pr2 | 4$e$ | 0 | 0.24890(2) | 1/4 | 5.36(9) | 5.2(1) | 4.4(1) | 6.4(1) | 0 | 0.8(1) | 0 |
| Na1 | 4$e$ | 0 | 0.9155(1) | 1/4 | 7.8(5) | 4.2(9) | 5.5(9) | 14(1) | 0 | 2.1(8) | 0 |
| Na2 | 4$c$ | 1/4 | 1/4 | 0 | 15.1(4) | 18(1) | 17(1) | 10(1) | 0.4(7) | 3.8(8) | 5.2(8) |
| Na3 | 8$f$ | 0.2635(2) | 0.5786(1) | 0.0000(1) | 13.7(3) | 13.0(7) | 17.5(8) | 9.6(8) | 3.0(5) | -0.4(6) | -3.5(5) |
| O1 | 8$f$ | 0.1468(3) | 0.7319(2) | 0.1432(2) | 7.7(5) | 9(1) | 8.0(1) | 6(1) | 0.3(8) | -0.5(9) | 0.7(9) |
| O2 | 8$f$ | 0.0983(4) | 0.4157(2) | 0.1428(2) | 8.1(5) | 8(1) | 10(1) | 7(1) | -0.2(7) | 1.7(9) | 0.4(8) |
| O3 | 8$f$ | 0.1442(4) | 0.0975(2) | 0.1396(2) | 8.3(5) | 9(1) | 10(1) | 6(1) | -0.3(7) | 1.3(9) | -0.5(8) |

Table V. Comparison of the *y* parameter and occupancy of metal atoms in the honeycomb layer used in the various structural models of α-$Na_2PrO_3$. The notation of each atom is modified from that of refs.[32, 34] to match that of our single crystal study. The space group is *C2/c* in all the models and Pr1, Pr2, and Na1 are located at 4*e* (0, *y*, 1/4) site. In powder refinement, consideration of metal atom site-mixing leads to better refinement due to the presence of stacking fault (See the main text for the detail)[32, 34]. The estimated standard deviations of the refined values are shown in parentheses.

|  |  | This study |  | Hinatsu et al.[32] |
|---|---|---|---|---|
|  | Method | Single crystal | Powder Rietveld refinement | Powder Rietveld refinement |
| Pr1 | *y* | 0.58165(2) | 0.58165 | 0.5646(9) |
|  | Occupancy of Pr/Na | 1/0 | 0.774(3)/0.226(3) | 0.34(1)/0.66(1) |
| Pr2 | *y* | 0.24890(2) | 0.2489 | 0.2494(6) |
|  | Occupancy of Pr/Na | 1/0 | 0.774(3)/0.226(3) | 1/0 |
| Na1 | *y* | 0.9155(1) | 0.9155 | 0.9181(7) |
|  | Occupancy of Na/Pr | 1/0 | 0.547(6)/0.453(4) | 0.34(1)/0.66(1) |

**Table VI**. Crystallographic data and the results of structural refinement of $\beta$-Na$_2$PrO$_3$ from single-crystal X-ray data at 293(2) K (reproduced from Ref.[38]). We use origin choice 2 for the *Fddd* space group (with inversion at origin). Peaks with $I > 2\sigma(I)$ are defined as observed ones. Atomic coordinates, and equivalent isotropic $U_{iso}$ and anisotropic $U_{ij}$ displacement parameters (in units of 10$^{-3}$Å$^2$) with the estimated standard deviations in parentheses.

| Formula | Na$_2$PrO$_3$ |
|---|---|
| Space group | *Fddd* |
| *a* (Å) | 6.7641(2) |
| *b* (Å) | 9.7866(4) |
| *c* (Å) | 20.5517(6) |
| *V* (Å$^3$) | 1360.47(8) |
| *Z* | 16 |
| $\mu$ (mm$^{-1}$) | 14.383 |
| $F_{000}$ | 1680 |
| $R_{p,obs}$ | 1.85 |
| $R_{p,all}$ | 3.09 |
| $R_{wp,obs}$ | 3.25 |
| $R_{wp,all}$ | 3.72 |
| $R_{int}$ | 5.88 |
| Index range | -9 < h < 9 |
|  | -13 < k < 12 |
|  | -27 < l < 26 |
| $N_{measured}$ | 4958 |
| $N_{unique}$ | 464 |
| $N_{obs}$ | 325 |
| $N_{param}$ | 31 |
| Crystal size ($\mu$m$^3$) | 170×95×31 |
| extinction coefficient $\xi$ | 8.0(3) × 10$^{-4}$ |

| Atom | Wyckoff | x | y | z | $U_{iso}$ | $U_{11}$ | $U_{22}$ | $U_{33}$ | $U_{12}$ | $U_{13}$ | $U_{23}$ |
|---|---|---|---|---|---|---|---|---|---|---|---|
| Pr1 | 16g | 1/8 | 1/8 | 0.70879(1) | 6.4(1) | 7.9(2) | 6.8(2) | 4.7(2) | 0.7(2) | 0 | 0 |
| Na1 | 16g | 1/8 | 1/8 | 0.0463(1) | 15.6(5) | 8(1) | 25(1) | 13(1) | 2(2) | 0 | 0 |
| Na2 | 16g | 1/8 | 1/8 | 0.8796(1) | 11.3(5) | 15(1) | 10(1) | 9(1) | 5(1) | 0 | 0 |
| O1 | 16e | 0.8400(5) | 1/8 | 1/8 | 10.0(9) | 11(2) | 10(2) | 9(2) | 0 | 0 | 0.5(18) |
| O2 | 32h | 0.6384(5) | 0.3522(3) | 0.0336(1) | 9.7(6) | 10(2) | 11(1) | 9(1) | 1(2) | 0.5(12) | -0.9(11) |

**Table VII**. Bond lengths and bond angles of the nearest Pr-O-Pr triangle for $\alpha$- and $\beta$-Na$_2$PrO$_3$ at 293 K with estimated standard deviations in parentheses.

| Compound | Pr-O-Pr | $d_{Pr-Pr}$ (Å) | $\theta_{Pr-O-Pr}$ (°) |
|---|---|---|---|
| $\alpha$-Na$_2$PrO$_3$ | Pr1-O1-Pr2 | 3.4434(2) | 100.93(6) |
|  | Pr1-O2-Pr2 | 3.4397(4) | 100.6(1) |
|  | Pr1-O3-Pr2 | 3.4434(2) | 99.33(8) |
| $\beta$-Na$_2$PrO$_3$ | Pr1-O1-Pr1 | 3.4442(6) | 99.6(1) |
|  | Pr1-O2-Pr1 | 3.4226(3) | 100.2(1) |